\documentstyle[preprint,aps]{revtex}
\draft
\preprint{SNUTP 99-037,\ KIAS-P99061}
\tightenlines
\begin{document}
\title{\Large\bf Model-dependent Axion as Quintessence
with Almost Massless Hidden Sector Quarks}
\author{Jihn E. Kim\footnote{e-mail: jekim@phyp.snu.ac.kr} } 
\address{
Department of Physics andCenter for Theoretical Physics, 
Seoul National University,
Seoul 151-742, Korea, and\\
School of Physics, Korea Institute for Advanced Study, 
207-43 Cheongryangri-dong, Dongdaemun-ku, Seoul 130-012, Korea
}
\maketitle

\begin{abstract} 
A pseudo-Goldstone boson for {\it quintessence} is known to require 
the decay constant
$F_q\sim M_P$ and the height of the potential $(0.003\ {\rm eV})^4$,
and hence the pseudo-Goldstone boson mass of order $10^{-33}$~eV.
The model-dependent axion in some superstring models is suggested
as a good candidate for the quintessence. 
We realize this idea with a hidden-sector gauge group 
$SU(3)_H\times U(1)_H$. The hidden-sector instanton contribution
to the quintessence potential is suppressed by the almost massless
hidden-sector quarks. The model-independent axion is
the QCD axion required for the strong CP solution.    
\end{abstract}

\keywords{[Keywords: axion, quintessence, cosmological constant, 
supergravity, hidden sector]}

\pacs{PACS: 98.80.Cq, 14.80.Mz, 12.60.Jv, 11.30.Rd}

\newpage

Various mass scales in physics have been the most unambiguous hints toward
new physics. For example, the weak scale $\sim 250$~GeV has led to 
the construction of the standard model. 
The Einstein-Hilbert action with the scale parameter
$M_P=2.44\times 10^{18}$~GeV gives the long range gravitational
interaction. From the huge ratio of these
two known scales, technicolor and supersymmetry solutions
of the hierarchy problem were proposed~\cite{hierarchy}, which
triggered the main research activities of high energy physics 
in the last twenty years.  In addition there exists a theoretically
favored (but still experimentally unproven) axion scale
at the window $10^{9-12}$~GeV~\cite{revaxion} which awaits to
be uncovered in the near future~\cite{search}. Also,   
there exists the scale $\mu$ in supersymmetric models
which is in principle an independent parameter~\cite{mu}.

In addition, another very important scale 
revealed gradually in the recent years from the shadow
is the vacuum energy density
or the cosmological constant.  Since the Einstein gravity without the
cosmological term was so successful, it has been the theoretical
prejudice that it must vanish. Experimentally, one could have given at most
an upper bound on the cosmological constant before 1998.
But with the first round observation
on the negative decceleration parameter in 1998~\cite{perm}, 
which has to be confirmed
with more data, the hypothesis of a positive but a small nonzero 
cosmological constant has attracted a great 
deal of attention~\cite{freeman}. 
If confirmed, it introduces a new physical scale of order
\begin{equation}
\Lambda_{c.c.}=(0.003\ {\rm eV})^4\simeq 10^{-46}\ {\rm GeV}^4.
\end{equation}
This scale for the cosmological constant is so small that it must
come from a ratio of a hierarchical number or from an exponentially
suppressed number. From the perturbative analysis, one usually
obtains a hierarchical ratio. If so, a plausible ratio is
\begin{equation}
\frac{v^7_{ew}}{M_P^3}\sim 4.2\times 10^{-39}~{\rm GeV}^4\ ,\ \ 
\frac{v^8_{ew}}{M_P^4}\sim 4.3\times 10^{-55}~{\rm GeV}^4
\end{equation}
where $v_{ew}$ is the known electroweak scale.
The term suppressed by $M_P^4$ gives a too small number responsible
for the nonzero cosmological constant since the possible couplings
would reduce the number further. The term suppressed by $M_P^3$ is
a bit large but the couplings multiplied would lower the number
further, and can lead to a desired vacuum energy of order
$10^{-46}\ {\rm GeV}^4$.   In this scenario, the vacuum energy  
suppressed by $M_P^3$ can interpret the correct magnitude of
the negative decceleration parameter~\cite{kimq}. 

On the other hand, there exists another plausible scale for
supersymmetry breaking, the intermediate scale at 
$10^{10-13}$~GeV~\cite{nilles}.
In Eq.~(2), we did not consider this scale under the assumption 
that the real physical parameters at the observable sector
are provided by the mass splitting of superpartners. In this
theory, the vacuum energy and its one-loop correction are of order
the intermediate scale~\cite{chkini}, but in a theory with the vanishing
cosmological constant following the philosophy of Ref.~\cite{kimq},  
our starting point is that the vacuum energy is zero at the minimum of the 
potential.  However, this intermediate scale must be scrutinized in 
specific quintessence models.

The best candidate for quintessence is a pseudo-Goldstone
boson~\cite{freeman,kimq,choi}.\footnote{For more quintessence idea, see
Ref.~\cite{quintref}.} 
In string models, usually the string axions
belong to this category~\cite{kimq,choi}. The key point for
this argument is that in a specific direction of the parameter
space the potential is almost flat to realize the extremely
shallow quintessence potential. Namely, the fundamental theory
must have a spontaneously broken global symmetry with a very tiny
explicit breaking term of order $1/M_P^3$~\cite{kimq}. 

In supergravity, the supersymmetry breaking is introduced by
the confining force in the hidden sector~\cite{nilles,din}.
As usual, we introduce the hidden sector 
as appearing at $10^{10-13}$~GeV which contains a confining
gauge group.

Even though the model presented in Ref.~\cite{kimq}, with
a hidden sector $SU(2)_H$, 
nicely introduces the quintessence vacuum energy, the 
model itself destroys the successful gauge coupling unification 
due to the introduction of the extra $SU(2)_W$ triplet fields
in the model.
The $SU(2)_W$ gauge group has been used in Ref.~\cite{kimq} 
to protect the hidden-sector (h-sector) quarks from obtaining mass. 
Toward the gauge coupling unification, a gauge group needed
to protect the h-sector quark from obtaining mass might be
restricted to $U(1)$'s so that the difference of $SU(3)_c$
and $SU(2)_W$ $\beta$ functions is not changed from the
minimal supersymmetric standard model(MSSM). The $\beta$ function
for $U(1)_Y$ can also be unchanged if the h-sector quarks carry
zero electromagnetic charge.
Therefore, we must decouple the hidden sector from the observable
sector. 

In this paper, we try to build a quintessence model purely from
the theory of the hidden sector decoupled from the observable sector.
The hidden sector nonabelian group can be semi-simple, but we 
will restrict to the simple group $SU(3)_H$
as an example. The group
$SU(2)$ is smaller, but starting with the universal coupling constant
at the Planck scale, $SU(2)$ is not large enough to be strong at
the intermediate scale. With $SU(3)_H$, one can introduce a small
number of h-quarks than in QCD and can make the coupling strong
at the intermediate scale.\footnote{If needed, one can introduce
$SU(4)_H$ not changing our conclusion.} We will also introduce  
a $U(1)$ factor gauge group, $U(1)_H$, and 
the h-sector gauge group is assumed to be 
\begin{equation}
SU(3)_H\times U(1)_H.
\end{equation}
If the extra superfields do not carry any quantum number of
$SU(3)_c\times SU(2)_W\times U(1)_Y$, the h-sector is a real
hidden sector from the observable sector. Its effect is
transmitted to the observable sector only by gravity.
We introduce four h-sector chiral quarks(h-quarks) and 
three singlets whose quantum numbers are shown in Table~1.

\vskip 0.5cm
\centerline{Table~1. The $SU(3)_H$ and $U(1)_H$ 
quantum numbers of superfields.}
\vskip 0.5cm
\begin{center}
\begin{tabular}{|c|c|c|c|c|c|c|c|}
\hline
&&&&&&&\\
&\ \ $Q_1$\ \ &\ \ $Q_2$\ \ &\ \ $\bar Q_3$\ \ &\ \ $\bar Q_4$\ \ &\ \  
$S_1$\ \ &\ \ $S_2$\ \ &\ \ $S_0$\ \ \\
&&&&&&&\\
\ \ $SU(3)_H$\ \  &${\bf 3}$&${\bf 3}$&${\bf 3}^*$
&${\bf 3}^*$&{\bf 1}&{\bf 1}&{\bf 1}\\
&&&&&&&\\
\ \ $U(1)_H$\ \  &4&--4&0&0&1&--1&0\\
&&&&&&&\\
\hline
\end{tabular}
\end{center}
\vskip 0.5cm
The renormalizable terms for the singlets are
\begin{equation}
W=f_0S_0^3+f_1S_1S_2S_0-\mu S_1S_2-\mu^\prime S_0^2+
\mu^{\prime\prime 2}S_0.
\end{equation}
The vanishing $U(1)_H$ D-term sets $|\langle S_1\rangle|=|\langle
S_2\rangle|$. Vanishing of the F-term gives
\begin{equation}
\tilde v(-\mu+f_1S_0)=0,\ \ 
3f_0S_0^2+ f_1S_1S_2-2\mu^\prime S_0+\mu^{\prime\prime 2}=0,
\end{equation} 
where we have the usual $\mu$ and $\mu^\prime$ terms, and 
$|\langle S_1\rangle|=|\langle S_2\rangle|=\tilde v$.
$\mu, \mu^\prime$ and $\mu^{\prime\prime}$ are 
assumed to be at TeV scale, for which there can
be many solutions~\cite{mu}. These solutions assume
some kind of symmetry such as $U(1)_{PQ}$, $U(1)_R$, or
others. If the symmetry is global, one expects a Goldstone
boson when it is broken. In this paper, however, we will try to introduce
a global symmetry from a gauge symmetry and hence treat these mass
parameters $\mu,\mu^\prime$ and $\mu^{\prime\prime}$ just as inputs. 
We will comment more on the $\mu$ parameter later. 

Let $\langle S_1\rangle
=e^{i(\alpha+\beta)}\tilde v$ and $\langle S_2\rangle 
=e^{i(\alpha-\beta)}\tilde v$. Then, we obtain
\begin{equation}
\tilde v=\frac{e^{-i\alpha}\mu}{f_1}\sqrt{-\left(3\frac{f_0}{f_1}
-2\frac{\mu^\prime}{\mu}+f_1\frac{\mu^{\prime\prime 2}}{\mu^2}\right)}.
\end{equation}
Thus the $U(1)_H$ symmetry is broken at a TeV scale and we
expect a neutral gauge boson at this scale. Since this hidden
sector $U(1)_H$ interacts with the observable sector only through
gravity, it is not expected for this extra $U(1)_H$ gauge boson
to be found at high energy accelerators. 

The most dominant $SU(3)_H\times U(1)_H$ invariant couplings 
are
\begin{equation}
\frac{1}{M_P^3} Q_1\bar Q_3 S_2^4,\ \ 
\frac{1}{M_P^3} Q_1\bar Q_4 S_2^4,\ \ 
\frac{1}{M_P^3} Q_2\bar Q_3 S_1^4,\ \ 
\frac{1}{M_P^3} Q_2\bar Q_4 S_1^4,
\end{equation}
which can lead to the desired vacuum energy since it is suppressed
by $M_P^3$. The terms given in Eq.~(7)
break the following chiral symmetry
\begin{equation}
\{Q_1, Q_2, \bar Q_3, \bar Q_4\}\rightarrow e^{i\alpha}\{
Q_1, Q_2, \bar Q_3, \bar Q_4\}.
\end{equation}
The $S_i$ fields are invariant under the above chiral transformation.
The h-quarks would obtain mass of order $\tilde v^4/M_P^3\sim 
2.7\times 10^{-46}$~GeV which leads to the almost zero
h-sector instanton potential, 
\begin{equation}
\sim \frac{m_Q^2 M^3_{\rm h-gluino}}{\Lambda_h}\sim 10^{-98}~{\rm GeV}^4
\nonumber
\end{equation}
for $M_{h-gluino}\sim 100$~GeV, $M_Q\sim S_i^4/M_P^3\ (i=1,2)$
and $\Lambda_h\sim 10^{13}$~GeV. 
Therefore, the instanton potential is not important, compared to the 
above explicit chiral symmetry breaking terms given in Eq.~(7). The
h-squarks (the superpartner of h-quark) are assumed to have no
condensation. This point is essential for interpreting a model-dependent
axion $a_{MD}$ as the quintessence.

First we note the following.
If there is a hierarchy between two scales for two pseudoscalar 
bosons both of which
couple to two independent explicit symmetry breaking terms, then
one can easily estimate the decay constants and masses of the
bosons due to the {\it inverted order decay constant theorem}: 
{\it The smaller decay
constant corresponds to the larger explicit symmetry breaking 
scale, and the larger decay constant corresponds to the smaller
explicit symmetry breaking scale}~\cite{staxion,ckn,kimq}. 

To show this, let us consider two Goldstone bosons, $a_1$ and $a_2$, 
with decay constants $F_1$ and $F_2$, respectively, arising from
the spontaneous breaking scales $F_1$
and $F_2$ of two global symmetries.
Without explicit symmetry breaking terms, there will be no potential
depending on $a_1$ and $a_2$, due to the Goldstone theorem.
If there exist explicit symmetry breaking terms, there result
$a_1$- and $a_2$- dependent potentials which must be periodic 
functions of $a_1$ and $a_2$. Since the original global symmetry can
be called phase symmetries, the fields must return to itself after
shifts of $a_1\rightarrow a_1+2\pi F_1/\alpha_1$ or $a_2\rightarrow 
a_2+2\pi F_2/\alpha_2$ where $\alpha_1$ and $\alpha_2$ are determined
by the quantum numbers of the fields breaking $U(1)$ global 
symmetries. Therefore, if there exist two explicit 
symmetry breaking terms, then the potential is of the following form,
$$
V\sim -V_1\cos\left(\alpha_1\frac{a_1}{F_1}+\alpha_2\frac{a_2}{F_2}
\right)-V_2\cos\left(\beta_1\frac{a_1}{F_1}+\beta_2\frac{a_2}{F_2}
+{\rm constant}\right)
$$
where the hierarchical explicit symmetry breaking scales are parametrized
by $V_1$ and $V_2$, and both $\alpha_1$ and $\alpha_2$ are 
nonzero(i.e. two disconnected sector is not considered). 
To give masses to both bosons, we require $\alpha_1/\alpha_2\ne\beta_1/
\beta_2$. For a hierarchy, $V_1\gg V_2$, one Goldstone boson determined
by the dominant symmetry breaking scale
is approximately,
$$
\alpha_1\frac{a_1}{F_1}+\alpha_2\frac{a_2}{F_2}\propto
{a^\prime\over F^\prime}
$$ 
where 
$$
F^\prime={F_1F_2\over\sqrt{\alpha_1^2F_2^2+\alpha_2^2F_1^2}}.
$$
$F^\prime$ is the decay constant corresponding to the mass
eigenstate $a^\prime$. If there is a hierarchy among the decay
constants, say $F_1\gg F_2$, then we obtain $F^\prime\simeq
F_2/\alpha_2$, i.e. the decay constant corresponding to the
larger explicit symmetry breaking scale $V_1$ 
is the smaller (i.e. $F_2$) of $F_1$
and $F_2$, leading to the inverted order theorem.
Orthogonalization of the mass matrix leads to the other
mass eigenstate whose decay constant is the larger scale
$F_1$ and explicit symmetry breaking scale is the smaller one
$V_2$.

There are two kinds of axions in superstring models.
The model-independent axion is present in any superstring models.
It is basically the second rank antisymmetric tensor field $B_{\mu\nu}
(\mu,\nu= {\rm 4D\ indices})$
after compactification. The model-dependent axion couples to the fields
universally. In addition, there exist model-dependent axions $B_{ij}\ 
(i,j= {\rm internal\ indices})$
whose properties depend on how the model is compactified. The numbers of
model-independent and model-dependent axions are the zeroth and the
second Betti numbers, respectively. 

One very well-known example of the above theorem 
is the $\eta$--axion mixing.
The chiral symmetry breaking of QCD introduces the QCD scales
$\sim$~1~GeV and 100~MeV as the explicit symmetry breaking scales 
for $\eta$ and axion, respectively. The decay constants
of $\eta$-meson and axion are of order 1 GeV and $F_a\sim 
10^{9-12}$~GeV~\cite{revaxion}, respectively. Due to the above theorem, 
$\eta$ obtains a mass of order 1 GeV, and axion obtains a mass of
order $(0.1{\rm \ GeV})^2/F_a$. Another example is the axion decay
constant calculated in Ref.~\cite{kimq,gkn}.

In the present case, however, we have four Goldstone bosons,
the model-independent axion $a_{MI}$, a model dependent axion
$a_{MD}$, the pseudoscalar $a_{h-gl}$ arising from the h-sector gluino
condensation of order $\Lambda_h^3\sim (10^{13}{\rm \ 
GeV})^3$, and the pseudoscalar $a_{h-qu}$ arising from the h-quark
condensation below the electroweak scale. 
\footnote{If the fermion 
bilinears condense, then supersymmetry is broken. In this paper, we
assume supersymmetry below the h-sector scale, and hence
the h-quark condensation scale is assumed to be of order 
the mass splitting between superpartners.}

Therefore, in the present case 
one cannot state as simply as for the case of two Goldstone bosons.
Toward an explicit discussion,
we will choose the decay constant scales for $a_{MI}$ and 
$a_{MD}$ as $F_a\sim 10^{12}$~GeV and $
F_q\sim M_P$, respectively~\cite{kimq}. 
The decay constant for $a_{h-gl}$ is $f_{h-gl}\Lambda_h$, and 
the decay constant for $a_{h-qu}$ is about the electroweak scale,
$\langle {\rm h-quark\cdot h-quark}\rangle\sim (f_{h-qu}v)^3$
where $v\simeq 250$~GeV. Here, $f$'s are dimensionless parameters.

The explicit breaking scales giving Goldstone bosons tiny masses 
are the h-sector gluino condensation $\Lambda_h$, 
the QCD scale $\Lambda_{\rm QCD}$, 
$\epsilon_{h-qu}\sim v^{7/4}/
M_P^{3/4}$ from Eq.~(7), $\epsilon_\Lambda$ which arises
from the h-quark mass suppressed h-sector instanton potential,
and $f_{h-gl}^{3/2}\Lambda_h^{3/2}/M_P^{1/2}$. Numerically, 
$\epsilon_\Lambda$ is negligible compared to the others, and hence 
we will neglect $\epsilon_\Lambda$.   
Among these, the largest explicit breaking scale is
$\Lambda_h$. The Goldstone boson affected by $\Lambda_h$ is 
$a_{h-gl}$. Thus, $a_{h-gl}$ obtains a mass of order $\Lambda_h$. 
Among the remaining scales, the largest explicit breaking scale
is $\Lambda_{QCD}$.
The Goldstone boson $a_{h-qu}$ corresponding to the h-quark is 
separated from QCD. Thus, among the other two, 
$a_{MI}$ and $a_{MD}$, where we choose
their decay constants at the scales $10^{12}$~GeV and $M_P$, 
respectively, $a_{MI}$ corresponds to $\Lambda_{\rm QCD}$ and
$a_{MD}$ corresponds to $M_P$ according to the {\it inverted order 
decay constant theorem.} Thus, the very light axion for the strong
CP solution can be realized by $a_{MI}$ with an appropriate
$F_a\sim 10^{12}$~GeV~\cite{kimq,kim}. 
Now the shallow potential for the remaining $a_{MD}$ is compared
to the other explicit breaking terms and we can consider $2\times 2$
mass matrix of the remaining Goldstone bosons, $a_{MD}\simeq a_q$
and $a_{h-qu}$.
These two pseudoscalars have the hidden sector interaction. 
The two decay constants have a
hierarchy $F_q\sim M_P\gg f_{h-qu}v$. Thus, again from
the {\it inverted order decay constant theorem},
$a_{h-qu}$ with the decay constant $\sim v$ corresponds to
the explicit breaking scale $\epsilon_{h-qu}$ and $a_q$
corresponds to the explicit breaking scale 
$(10^{-46}\ {\rm GeV}^4)^{1/4}\simeq 10^{-11.5}\ {\rm GeV}$. 
Thus the mass of the Goldstone boson $a_{h-qu}$
is about $\epsilon_{h-qu}^2/f_{h-qu}v\sim 2.6\times 10^{-22}\ {\rm
GeV}/f_{h-qu}$, and  the remaining $a_q$ which we interpret as
quintessence has the decay constant $F_q\sim M_P$ 
and mass $\sim 10^{-33}$~eV.
Since the model-dependent axion which arises from the compactification
scheme is interpreted as quintessence,
it is possible to expect $F_q\sim M_P$ in some compactification
models.

Now we proceed to discuss the couplings of the quintessence to
matter fields. Let us assume that the quintessence is
a model-dependent axion so that we use 
$a_{MD}\simeq a_q$ with the decay constant $F_{MD}\simeq F_q$.
Then, we can imagine the couplings of $a_{MD}\simeq a_q$, being moduli, 
\begin{equation}
\frac{1}{M_P^3}e^{ia_{MD}/F_q} Q_1\bar Q_3 S_2^4, \cdots. 
\end{equation}
Then the terms given in Eq.~(7) is invariant under the 
transformation (8) with the shift of $a_{MD}$
\begin{equation}
a_{MD}\rightarrow a_{MD}-\alpha F_{MD}.
\end{equation} 
If the first term in Eq.~(10) is the only allowed coupling, 
then the supersymmetric F-terms do not generate a potential for
$a_{MD}$. However, there can exist other terms denoted by dots.
In this case, the supersymmetric terms can generate $a_{MD}$
dependent potential; but these are suppressed by $M_P^6$,
viz. $V\sim \sum_i|\partial W/\partial\phi_i|^2$. 
With a broken supersymmetry, the soft terms, i.e. $m_{3/2}\cdot 
(Eq.~(10))$, can give a potential depending on $a_q$.  
This is a situation we imagine to be realized for a natual
quintessence.
The vacuum energy generated by $\langle Q_1\bar Q_3 S_2^4 \rangle$
breaks the above shift symmetry, and the Goldstone boson corresponding
to the shift symmetry obtains a mass of order $(\langle V\rangle/F_q^2
)^{1/2}\sim 10^{-33}$~eV. For this argument to be valid, the
superpotential terms are restricted by some kind of symmetry (using global
and discrete symmetries) so that superpotential terms 
with $D\le 5$ are forbidden.
 
But in string models, the global symmetries are badly broken by the
world-sheet instanton effect~\cite{ww,dsww}. 
The coupling of the model-dependent axion is given by
\begin{equation}
W\sim \Phi_1\Phi_2\Phi_3(1+\epsilon e^{-R/M_P})
\end{equation} 
where $\epsilon$ denotes the importance of the world-sheet contribution,
$R=r-ia_{MD}$ is the chiral field containing $a_{MD}$, 
and $\Phi_1\Phi_2\Phi_3$ represents a generic superpotential term.
As discussed in the previous paragraph, coupling of 
the model-dependent axion to superpotential terms with 
$D\le 5$ can be too large
for a successful identification of $a_{MD}$ as $a_q$.
But there is a caveat in this argument.
The dangerous $a_{MD}$ coupling in Eq.~(12) for it to be
quintessence arises if
the vacuum expectation value $\langle\Phi^3\rangle$ is
much larger than $(0.003\ {\rm eV})^4$. 

If the terms in the superpotential 
do not generate nonvanishing VEV's, then
the model-dependent axion coupling is negligible. 
{}From the superpotential of the type (12), one obtains
soft terms and fermionic terms of the following
form,
\begin{eqnarray}
&V_{\rm soft}\sim m_{3/2}\Phi_1\Phi_2\Phi_3
(3+3\epsilon e^{-R/M_P}+\epsilon e^{-2R/M_P})+{\rm h.c.}\\
&{\cal -L}_Y\sim -\psi_{\Phi_1} \psi_{\Phi_2}
\Phi_3 (1+\epsilon e^{-R/M_P})+{\rm h.c.}+({\rm permutations})
+({\rm axino\ terms}),
\end{eqnarray}
where $\psi_{\Phi_i}$ field is the fermionic superpartner of $\Phi_i$.

Interactions are obtained from the superpotential (12) where we suppressed
couplings for simplicity.  The potential 
\begin{equation}
V=\sum_{\phi_i=\Phi_1,\Phi_2,\Phi_3,R}\left|\frac{\partial W} 
{\partial \phi_i}\right|^2
\end{equation}
includes the $a_{MD}$-dependent terms of the following form,
\begin{equation}
\left|\Phi_1\Phi_2 (1+\epsilon e^{-R/M_P})\right|^2
+({\rm permutations}),
\end{equation} 
from which we extract dominant supersymmetric $a_{MD}$-dependent 
couplings,
\begin{equation}
2\epsilon e^{-r/M_P}
|\Phi_1\Phi_2|^2\cos\frac{a_{MD}}{M_P} +({\rm permutations})
\end{equation}
where we shifted $a_{MD}$ so that $\epsilon$ becomes real.

From the above example, one can see that the most 
dangerous terms can come from superpotentials
constructed from $SU(3)\times SU(2)\times U(1)$ singlet fields.
If all the singlets involved in the couplings,viz. Eqs.~(15)
and (16), obtain superheavy
vacuum expectation values, our argument for the quintessence
cannot be realized. Therefore, we assume that superpotentials
constructed with singlet fields do not give nonvanishing vacuum
energy, namely at least one scalar component of $\Phi_i$ 
in Eq.~(13) has the vanishing VEV. For Eq.~(16) not to
contribute, at most one out of $\Phi_1,\Phi_2$ and $\Phi_3$
can have a nonvanishing VEV. 
Note that the fermionic components of the singlets do not condense
and Eq.~(14) is not particularly important for the singlets.

\def\half{\frac{1}{2}}
But Eq.~(14) can be important for the 
fields of the minimal supersymmetric standard model(MSSM). 
In the MSSM let us first consider $V_{\rm soft}$.
The A-terms of the MSSM Yukawa couplings in the charge
and color conserving vacuum do not generate a $a_{MD}$ dependent
potential. Among the MSSM couplings, therefore, the relevant term is
the B-term from the $\mu$-term~\cite{mu} since $\langle H_1H_2\rangle\ne 0$. 
If the $\mu$ term is corrected as above
by the world-sheet instanton effect, then the model-dependent axion
cannot be identified as the quintessence. However, the argument
given in Ref.~\cite{dsww} does not apply to the $\mu$-term.\footnote{
In Ref.~\cite{dsww}, $\Phi_1\Phi_2\Phi_3 e^{-R}$ coupling was obtained
by considering the expectation value of the dilaton vertex operator
$V_B$ between the fermionic states, e.g. $\langle \half,\half,\half,
-\half|V_B|-\half,-\half,-\half,\half\rangle$. For the $\mu$ term, however,
one needs to calculate $\langle \half,\half,\half,-\half|1|
-\half,-\half,-\half,\half\rangle$ which seems to be vanishing.}

Next, the most dangerous term is the Yukawa couplings ${\cal L}_Y$.
The known quark condensation and VEV of the Higgs doublet fields
would generate an electroweak scale vacuum energy 
which is enormous compared to
the quintessence potential if it is coupled to $a_{MD}$.
But if the model-dependent axion potential to the MSSM fields is {\it 
universal}, then the Yukawa coupling is not dangerous. 
In this case, we can write the coupling as
\begin{equation}
-(1+\epsilon e^{-R})\cdot (\sum_{ij} f^{d}_{ij}\bar d^i q^j H_1
+\sum_{ij}f^{u}_{ij}\bar u^i q^j H_2)+{\rm h.c.}.
\end{equation}
where $i,j$ are the family indices.
At the minimum of the potential the vacuum energy is zero, i.e.
\begin{equation}
-\langle
\sum_{ij} f^{d}_{ij}\bar d^i q^j H_1
+\sum_{ij}f^{u}_{ij}\bar u^i q^j H_2
\rangle=0.
\end{equation}
Therefore, at the minimum the tree level $a_{MD}$ potential is
vanishing even if we consider the quark condensation. 

At the electroweak scale another relevant term in the superpotential
is the $\mu$ term,
$\mu H_1H_2$. If we treat $\mu$ as an input parameter, then
we can neglect its coupling to $a_{MD}$ through Eq.~(12), as 
commented above.  
But in some solutions of the $\mu$ problem, it is generated
through nonrenormalizable terms in which case we should consider
the coupling of the type Eq.~(12). Then it leads to a
too large potential for the model-dependent axion to be the quintessence.
For example, one may consider only cubic terms in the
superpotential $W_0$ inspired by string models~\cite{cm}.
Therefore, $\mu H_1H_2$ is absent in $W_0$. $W_0$ is assumed
to contain the singlet terms with the intermediate scale VEV's.  
Supergravity can contain nonrenormalizable terms 
$W_0H_1H_2/M_P^2$ which will give the desired magnitude
of $\mu$ for $\langle W_0\rangle\sim$~(intermediate scale)$^3$.
For this nonrenormalizable term, we expect that there exists
a world-sheet instanton correction which is not expected to
take the universal form Eq.~(12). In this case,
$a_{MD}$ cannot work as the quintessence. In the same vein, any 
solution of $\mu$ by the superpotential is thought to have 
the same problem.  However, if the world-sheet instanton 
correction to the nonrenormalizable potential also takes the universal 
form Eq.~(12), then the requirement of vanishing vacuum energy 
gives the $a_{MD}$ coupling negligible.

In conclusion, we investigated a model to interpret 
a model-dependent axion $a_{MD}$ 
as quintessence. For this idea to work, there must exist a
symmetry so that superpotential terms generating a vacuum energy 
of order $v_{ew}^6/M_P^2$ must be forbidden. We realize this
idea introducing a $SU(3)\times 
U(1)_H$ gauge symmetry in the hidden sector. 

\acknowledgments
I thank H. B. Kim for useful comments. I also
thank Korea Institute for Advanced Study for the hospitality
during my visit when this work is performed.
This work is supported in part by KOSEF,  MOE through
BSRI 98-2468, and Korea Research Foundation.


\begin{references}

\bibitem{hierarchy} S. Weinberg, {\it Implications of
dynamical symmetry breaking: an addendum, Phys. Rev.} 
{\bf D19} (1979) 1277;\\  
L. Susskind, {\it Dynamics of spontaneous symmetry breaking in the
Weinberg-Salam theory, Phys. Rev.} {\bf D20} (1979) 2619;\\ 
E. Witten, {\it
Dynamical breaking of supersymmetry, Nucl. Phys.} {\bf B188} (1981) 513.

\bibitem{revaxion} J. E. Kim, {\it 
Light pseudoscalars, particle physics and cosmology,
Phys. Rep.} {\bf 150} (1987) 1;\\
H. Y. Cheng, {\it 
The strong CP problem revisited, Phys. Rep.} {\bf 158} (1988) 1;\\ 
R. D. Peccei, {\it
The strong CP problem},
in {\it CP Violation}, ed. C. Jarlskog (World Scientific,
1989) p.503.

\bibitem{search} J. E. Kim, {\it Constraints on very light axions
from cavity experiments, Phys. Rev.} {\bf D58} (1998) 055006
[hep-ph/9802220];\\ 
C. Hagmann {\it et. al.}, {\it Results from a high-sensitivity
search for cosmic axions, Phys. Rev. Lett.} {\bf 80} (1998) 2043
[astro-ph/9801286].

\bibitem{mu} J. E. Kim and H. P. Nilles, {\it The $\mu$ problem
and the strong CP problem, Phys. Lett.} {\bf B138} (1984) 150;\\ 
G. F. Giudice and A. Masiero, {\it A natural solution to
the $\mu$ problem in supergravity theories Phys. Lett.} {\bf B206}
(1988) 480;\\ 
J. A. Casas and C. Munoz, {\it A natural solution to the
$\mu$ problem, Phys. Lett.} {\bf B306} (1993) 288 [hep-ph/9302227];\\
J. E. Kim and H. P. Nilles, {\it Symmetry principles toward
solutions of the $\mu$ problem, Mod. Phys. Lett.}
{\bf A9} (1994) 3575 [hep-ph/9406296].

\bibitem{perm}
S. Perlmutter {\it et al.}, {\it Measurements of the cosmological
parameters $\Omega$ and $\Lambda$ from the first 7 supernovae
at $z\ge 0.35$, Astrophys. J.}
{\bf 483} (1997) 565 [astro-ph/9608192];\\ 
The Supernova Cosmology Project,
{\it Cosmology from Type IA supernovae,
Bull. Am. Astron. Soc.} {\bf 5}
(1998) 1351 [astro-ph/9812473];\\ 
A. Riess, W. H. Press and R. P. Kirshner, {\it Using SN-IA light 
curve shapes to measure the Hubble constant}, submitted to
{\it Astrophys. J.} (1998);\\ 
S. Perlmutter and A. Riess, {\it Cosmological parameters
from supernovae: two group's results agree}, in COSMO-98,
ed. D. Caldwell (AIP Conference Proc. 478, Woodbury, N.Y., 1999),
p. 129 [Talk presented at COSMO-98,
Asilomar, Monterey, CA, Nov. 16--20, 1998].
  
\bibitem{freeman} J. Frieman, {\it Dynamical vacuum},
talk presented at COSMO-98,
Asilomar, California, Nov. 15--20, 1998 (unpublished).

\bibitem{kimq} J. E. Kim, {\it Axion and almost massless quark
as ingredients of quintessence, J. High Energy Phys.} 
{\bf 05} (1999) 022 [hep-ph/9811509].

\bibitem{nilles} H. P. Nilles, {\it Supersymmetry, supergravity
and particle physics, Phys. Rep.} {\bf 110} (1984) 1.

\bibitem{chkini} K. Choi, J. E. Kim and H. P. Nilles, 
{\it Cosmological constant and soft terms in supergravity,
Phys. Rev. Lett.} 
{\bf 73} (1994) 1758 [hep-ph/9404311].

\bibitem{choi} K. Choi, {\it String or M-theory axion as 
quintessence}, [hep-ph/9902292].

\bibitem{quintref} P. Binetruy, {\it Models of dynamical
supersymmetry breaking and quintessence, Phys. Rev.} {\bf D 60}
(1999) 063502 [hep-ph/9810553];\\ 
C. Kolda and D. H. Lyth, {\it Quintessential difficulties, Phys.
Lett.} {\bf B 458} (1999) 197 [hep-ph/9811375];\\ 
T. Chiba, {\it Quintessence, the gravitational constant and gravity,
Phys. Rev.} {\bf D 60} (1999) 083508 [gr-qc/9903094];\\
P. Brax and J. Martin, {\it Quintessence and supergravity,
Phys. Lett.} {\bf B 468} (1999) 40 [astro-ph/9905040];\\
A. Masiero, M. Pietroni and F. Rosati, {\it SUSY QCD and quintessence,
Phys. Rev.} {\bf D 61} (2000) 023504 [hep-ph/9905346];\\
M. C. Bento, O. Bertolami, {\it Compactification, vacuum energy
and quintessence, Gen. Rel. Grav.} {\bf 31} (1999) 1461 [gr-qc/9905075];\\
F. Perrotta, C. Baccigalupi and S. Matarrese, {\it Extended
quintessence, Phys. Rev.} {\bf D 61} (2000) 023507 [astro-ph/9906066].

\bibitem{din} J. P. Derendinger, L. Ibanez and H. P. Nilles,
{\it On the low-energy D=4, N=1 supergravity theory extracted
from the D=10, N=1 superstring, Phys. Lett.} {\bf B155} (1985) 65;\\ 
M. Dine, R. Rohm, N. Seiberg, and
E. Witten, {\it Gluino condensation in superstring
models, Phys. Lett.} {\bf B156} (1985) 55.

\bibitem{staxion} K. Choi and J. E. Kim, {\it Compactification and 
axions in $E_8\times E_8^\prime$ superstring models,
Phys. Lett.} {\bf B165} (1985) 71.

\bibitem{ckn} E. J. Chun, J. E. Kim and H. P. Nilles, {\it A natural
solution of the $\mu$ problem with a composite axion in the
hidden sector, Nucl. Phys.} {\bf B370} (1992) 105.

\bibitem{gkn} H. Georgi, J. E. Kim and H. P. Nilles, {\it
Hidden sector gaugino condensation and the
model independent axion,
Phys. Lett.} {\bf B437} (1998) 325 [hep-ph/9805503];\\  
J. E. Kim, {\it Superstring axion, gaugino condensation and discrete
symmetries}, in {\it PASCOS-98}, ed. P. Nath (World Scientific
Pub. Co., Singapore, 1999), p.361, 
talk presented at PASCO-98, Boston, March 22-29, 1998  [hep-ph/9807322].

\bibitem{kim} J. E. Kim, {\it Weak interaction singlet
and strong CP invariance, Phys. Rev. Lett.} {\bf 43} (1979) 103;\\
M. A. Shifman, V. I. Vainstein and V. I. Zakharov,
{\it Nucl. Phys.} {\bf B 166} (1980) 4933;\\
M. Dine, W. Fischler and M. Srednicki, {\it
A simple solution to the strong CP problem with a
harmless axion, Phys. Lett.} {\bf B104} (1981) 199;\\ 
A. Zhitniskii, {\it On possible suppression of the axion hadron
interactions}, (in Russian), {\it Sov. J. Nucl. Phys.} {\bf 31}
(1980) 260.

\bibitem{ww} X.-G. Wen and E. Witten, {\it World sheet instantons and
the Peccei-Quinn symmetry, Phys. Lett.} {\bf B166}
(1986) 397.

\bibitem{dsww} M. Dine, N. Seiberg, X.-G. Wen, and E. Witten, 
{\it Nonperturbative effects on the string world sheet, Nucl.
Phys.} {\bf B278} (1986) 769.

\bibitem{dyn} J. E. Kim, {\it A natural solution
of $\mu$ from the hidden sector, Phys. Lett.} {\bf B452} (1999) 255
[hep-ph/9901204].

\bibitem{cm} J. A. Casas and C. Munoz, {\it A natural solution to the
$\mu$ problem, Phys. Lett.} {\bf B306} (1999)
288 [hep-ph/9302227].

\end{references}
\end{document}